# Assessing the Impact of Space School UK

Daniel Robson[1,2], Henry Lau[1], Áine O'Brien[1,3], Lucy Williams[1], Ben Sutlieff[1,4], Heidi Thiemann[1,5], Louise McCaul[1], George Weaver[1] and Tracey Dickens[1]

[1]Space School & Senior Space School UK, School of Physics and Astronomy, University of Leicester, UK [2]School of Pharmacy, University of Nottingham, UK [3]School of Geographical and Earth Sciences, University of Glasgow, UK [4]Anton Pannekoek Institute for Astronomy, University of Amsterdam, Science Park 904, 1098 XH Amsterdam, The Netherlands [5]School of Physical Sciences, The Open University, UK

tp57@le.ac.uk, spaceschool@le.ac.uk

*Space School UK (SSUK) is a series of summer residential programmes for secondary school aged students, held at the University of Leicester over 3 weeks each year. Each programme involves space-related activities run by a team of mentors - currently including university students, graduates, teachers and young professionals associated with the space sector - all of whom attended SSUK as students themselves. It includes the 6-day Space School UK and the 8-day Senior Space School UK (collectively SSUK) which are for 13–15 and 16–18 year olds respectively.*

*This paper seeks to evaluate and present the benefits of SSUK to individuals who participate in the programme, organisations involved in the running of SSUK, and to highlight and promote these benefits to the wider UK and global space community.*

*We also address which facets of SSUK make for such an engaging and encouraging experience for the students, that are missing from students' traditional education. We seek to show how SSUK acts as an excellent example of how to bridge the gap between secondary and tertiary space education. Through an analysis of our alumni survey results, we show that attending SSUK has a significant impact on career choices and prepares students for Higher Education, regardless of background. Some groups, such as women, and those from non-selective schools, reported a higher impact in some of these areas than others. Metrics such as skills learned, goals achieved, alongside knowledge of careers and Higher Education possibilities are discussed for various demographics.*

I. INTRODUCTION

Space School UK and Senior Space School UK (collectively SSUK) are six- and eight-day fee-paying, non-selective, residential summer schools at the University of Leicester (UoL) for students aged 13–15 and 16–18 respectively. Industry sponsorship and involvement in SSUK has varied in the past, providing scholarship places to widen participation of the summer school. Most recently, approximately 20 students each year have been fully funded in this way. The schools are run by a team of mentors, all of whom are now alumni and study for/have studied a space-related degree.

The aims of SSUK are to educate and inspire young people through astronomy and space science, preparing students for space careers and giving them a taste of university life. Students participate in activities similar to those they would experience as STEM undergraduates, e.g. lectures, small group practical experiments and problem-solving workshops. Students also learn other space-related skills such as scuba diving, which gives them an experience of astronaut training and an understanding of the effects of microgravity on the human body. Students stay in university accommodation, except for one night of camping to observe the night sky, coinciding with the Perseids meteor shower for Senior Space School UK. Students also get hands on experience of practical astronomy through the use of UoL's ground-based telescopes.

Space School UK began in 1989 at Sevenoaks School, Kent. In 1990 it moved to Brunel University, then moved to its current home at UoL in 2000. To celebrate its 30-year anniversary, its alumni community were asked to complete an anonymous online survey regarding its impact. The authors used social media to contact SSUK alumni to ask them about their experience, their motivations for attending, and the impact on their lives and careers. Alumni were invited to complete the survey via posts on social media: Twitter, Facebook groups, and the SSUK Facebook page. These platforms have a combined ~2,500 followers, some of which will overlap and not all of whom are alumni (e.g. parents). The survey was available online for 2 weeks over July 2019. 144 people completed the survey (~6% of the estimated total alumni population of ~ 2400). When asked to rate whether alumni would recommend SSUK to young people interested in space, the average score was 4.96 out of 5, indicating that SSUK is regarded as a highly positive and worthwhile experience. This paper provides a detailed analysis of answers to selected questions from the survey, related to the goals outlined in Section II.A below.

II. METHODS

A. *Survey Goals*

Questions were chosen to determine how effective SSUK is at addressing the following goals:

1. Prepare students for the study of space-related subjects at university;
2. Inspire students to work and study space-related subjects beyond school and higher education;
3. Improve the knowledge, skills and understanding of students in space-related subjects.

The survey aimed to determine if there were trends for any particular demographics (e.g. age, gender, educational background, highest level of education) and how they felt SSUK addressed the goals listed above.

*B. Question Styles*

Demographic questions were asked with multiple choice or check box answers. Respondents were also asked several questions about the extent to which (on a scale of 1-5, where 1 is the lowest and 5 is the highest) they felt SSUK prepared them for higher education and their careers, particularly in STEM/space-related disciplines.

*C. Data Analysis*

Responses were coded by the authors in order to simplify data sets when free-form answers were given as an option, to allow for ease of trend identification. Where there were lower numbers of particular groups (e.g. some degree subjects), these were grouped together. An example of this was for those who identified their studies as either medicine, chemistry, biology, or earth science and geography-related, these subjects were coded as 'Other STEMM' (Science Technology Engineering Maths and Medicine). Pivot tables were then used to determine any trends in the 1-5 scale responses by different groups (such as subject studied, gender etc.). For any situation in which an "average score" for one of these 1-5 questions is being discussed, it is the population-weighted mean of that category.

In order to investigate the average scores given by different groups for survey questions, a standard error, $\alpha$, was calculated, using:

$$\alpha = \frac{\sigma}{\sqrt{N}}$$

whereby $\sigma$ is the standard deviation in the dataset and $N$ is the population size of that dataset.

III. RESULTS AND DISCUSSION

*A. Respondent Demographics*

Of the 144 respondents, 55% were female and 45% were male. The historic gender balance of SSUK is approximately 40% female and 60% male.

Before 2007, there was no separate Space School UK and Senior Space School UK, only a single school for all students aged 13-18. Only 14% of our respondents attended SSUK prior to its separation into two schools in 2007. For the last 4 years, Senior Space School UK's popularity grew such that there needed to be two senior schools each year, so the number of recent, younger, alumni is larger. Over half (52%) of respondents attended SSUK recently (2014-2018 inclusive) at least once and 71% of respondents had attended between 2009 and 2018. The young age of respondents is reflected in the highest level of education completed, with 36% of respondents citing A-Level or equivalent as their highest level of education, followed by Masters (30%), and Bachelors (19%). More female respondents have PhDs than male respondents (12.5% to 7.6%) whilst more men have Masters (33% to 26%).

Two thirds (66%) of attendees attended a comprehensive school or state-funded grammar school, and one third (31.7%) attended a private day or boarding school. The remaining 2.2% were home-schooled. As this portion was so low, answers from this group were omitted from our schooling-based analyses, due to the large associated uncertainty. 40% of respondents attended SSUK once, whilst 60% reported attending SSUK twice or more. 19.2% of respondents work, or have previously worked, as a mentor at SSUK (see limitations outlined in Section IV at the end of this manuscript).

*B. Influence on Higher Education Choices*

*1) Overview*

In this section, we explore the impact that attending SSUK had on the Higher Education (HE) paths of respondents. We follow this by discussing a breakdown of responses by qualification, gender and school background.

For the question 'If you went to/are at university/hold an offer, which of the following best describes the degree(s) you studied/intend to?', respondents could pick all that applied, the most common subject was physics-related with 61%, followed by 11% for engineering-related and 8% for maths-related (see Figure 1). Although these subjects are the most common routes for those enthusiastic about space, it is also encouraging that we find other subjects prevalent, with 5% studying Arts and Humanities and 4% studying Medicine and/or Veterinary Science. This implies that SSUK can be an enjoyable experience for those who are interested in space regardless of which degree they later pursue.

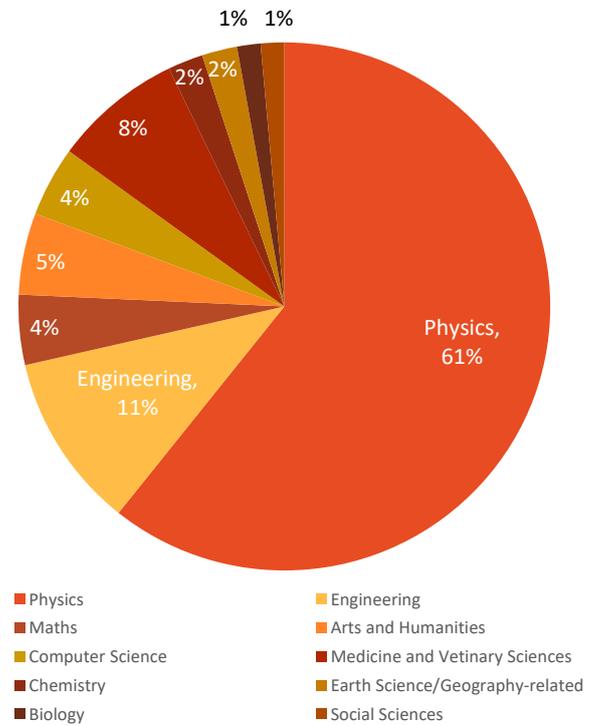

Fig. 1. The subjects studied by SSUK alumni survey respondents, when asked "If you went to/are at university/hold an offer, which of the following best describes the degree(s) you studied/intend to? (tick all that apply)"

4.1% of the UK student population in 2017/18 studied a physical science degree according to [1], which significantly contrasts with our respondents (61%). As well as physics being the most popular subject amongst our survey respondents, Figure 2 shows that those who studied Physics found that SSUK encouraged them to study it, more than any other subject. With an average score of $4.4 \pm 0.1$ for this question, it is evident that SSUK has strongly encouraged many students to pursue a physics-related degree.

Since 2000 the UoL School of Physics and Astronomy has hosted SSUK, also a factor in encouraging students to consider physics-related degrees. 89% of those surveyed had attended SSUK at the UoL. We also find that UoL is the most common university destination of respondents, at 13%. Furthermore, of the mentors (staff who provide both pastoral and educational support to students during the residential) who completed the survey, 79% have completed or are currently enrolled in a physics-related degree, and 30% of them have attended UoL. 52% of respondents felt that the mentors encouraged/prepared them for their studies/careers, which indicates that the alumni's choices are likely to have been influenced by both the subjects studied and university destinations of their mentors.

It is also encouraging to note that, regardless of degree subject, there was a very positive response to 'What impact did SSUK have on your opinions/plans of entering HE (higher education)?' and 'To what extent did Space School prepare you for HE?' (see Figure 2). This demonstrates that SSUK is a valuable experience for attendees to gain an insight into HE, regardless of whether they chose to study a STEM related degree. Furthermore, SSUK is successful in helping to bridge the gap between school and HE.

particularly as this is such an important discipline in the space sector. The lack of mentors who study computer science (7%) is also likely to have contributed to this.

When we look at how respondents with different levels of qualifications answered the question 'To what extent did SSUK encourage you to pick your degree subject?', we find that those with PhDs gave an average score of $4.3 \pm 0.2$, compared to $3.9 \pm 0.2$ for those with Master's degrees, $3.6 \pm 0.3$ for those with Bachelor's degrees and $3.9 \pm 0.1$ for those with A-level's or equivalent. Those who felt most influenced by SSUK to pick their degree subjects were those that went on to gain a PhD, indicating that SSUK has a lasting impact on those that felt strongly inspired by attending.

*2) Investigating influence on HE by gender*

When we look at which degree subjects female and male respondents studied, study, or intend to study, we find that pure physics-related degrees were more popular in women than men (43% to 39%), whilst more men chose engineering related degrees (16% to 4%). It is interesting to find that a higher proportion of female survey respondents studied pure physics than male respondents, given that, according to [1], 58% of physical science undergraduate students are male. Furthermore, grouping together all those who studied physics-related degrees (thereby including those who studied physics and other subjects) we find that 61% of female respondents studied physics, compared to 57% of male respondents. SSUK appears to have a positive effect on female students in encouraging them to study physics at university. This is supported by responses from female respondents who study physics giving an average rating of $4.5 \pm 0.1$ to the question ' To what extent did SSUK encourage you to pick your degree subject?' and $4.6 \pm 0.1$ to 'What impact did space school have on your opinions/plans of entering HE?'.

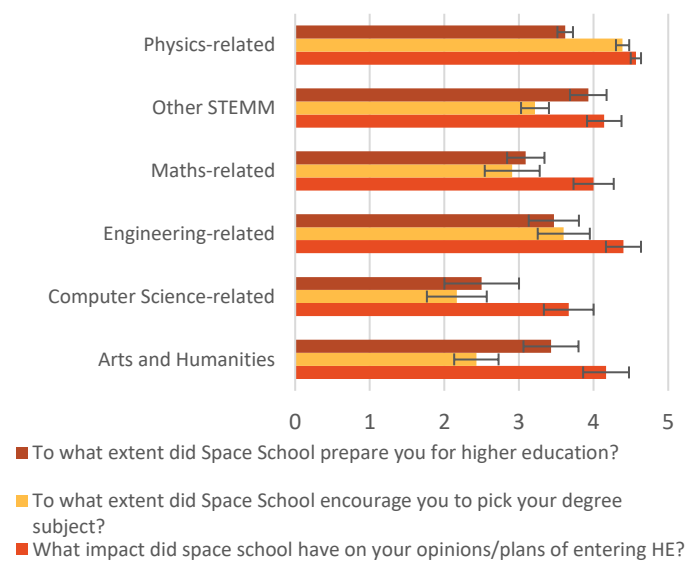

Fig. 2. Average scores to answers to HE questions in the SSUK alumni survey displayed by degree subject studied

We do, however, note that those who study computer science rated all three questions the lowest compared to other degree types. In the most recent years of SSUK, a programming workshop has been introduced, so we hope that this will have a positive impact on future alumni interested in computer science,

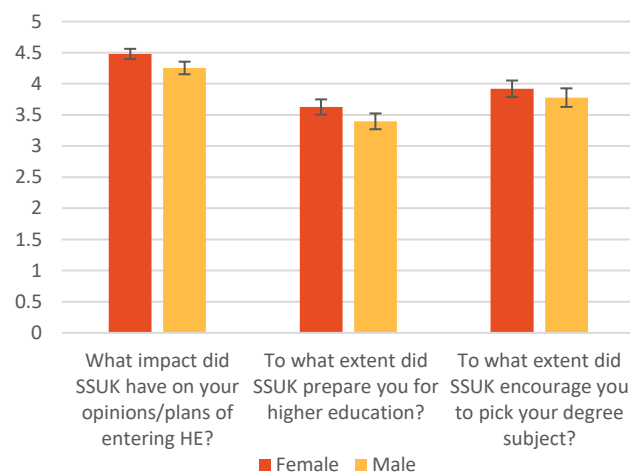

Fig. 3. Average scores of answers to HE questions in the SSUK alumni survey displayed by gender

When we look at the impact of SSUK on HE for male and female populations as a whole, we find some slight differences. Figure 3 demonstrates the average score for men and women for three HE-related questions. Although the average scores between male and female respondents agree within error, we do observe a trend, with female respondents answering more positively, on average, to all questions. This demonstrates that

SSUK holds a unique atmosphere of support and encouragement for girls into HE. One reason for this could be that 50% of mentors who responded are women, and 18% of female respondents answered 'mentors' to the question 'Which of these categories would you say has helped you most with your studies/career?', compared to 13% of male respondents. This highlights the importance of having female role models for girls, and how SSUK has provided this for many female alumni. This is explored further in part 4 of this Section.

*3) Investigating influence on HE by schooling type*

We can see from Figure 4 that there is variation in the impact SSUK had on HE choices with the type of schooling the respondents had.

For this study, those who answered that they either attended a private school, boarding school or state grammar school were grouped as having attended mostly selective schools. The authors recognise that not all boarding and private schools are selective with their intake. The scores for selective and non-selective schooling are largely consistent with one another, within error, indicating that SSUK adds value regarding HE choices, regardless of school background. However, the average scores for non-selective schooling are a little higher on average for all questions, indicating that those who attended comprehensive schools felt that they benefited the most from SSUK. The authors therefore encourage industry representatives (such as Lockheed Martin UK, Virgin Orbit and the Air League) to continue to sponsor places at SSUK under a remit of widening participation, to ensure those from lower socio-economic backgrounds can also access and benefit from SSUK.

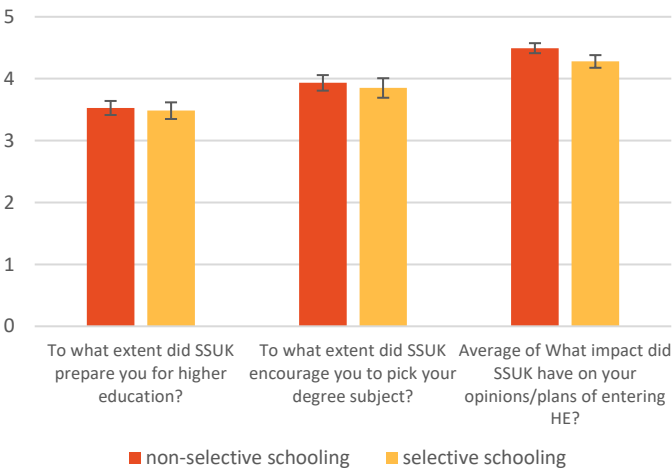

Fig. 4. Answers to HE questions in the SSUK alumni survey displayed by schooling type

*C. Influence on Career Choices*

*1) Overview*

The longevity of SSUK over the past 30 years enables this study to compare and review how alumni's attendance at SSUK has affected their career path. The opinions and insights of those who are in their mid-late careers are useful for not only the SSUK organisers but also for its students and the wider space sector. As evidenced by this report and below data, SSUK is of great importance for assisting alumni in achieving their career goals.

We find that 50% of survey respondents who have studied engineering stated they intended to work in the space sector, compared to just over 35% of physics-related graduates and students. An additional 21% of engineers currently or have previously worked in space. This is a largely positive results given the UK Space Agency's plan for the UK industry to be worth £30 billion by 2030 [4], whereby a larger technical workforce is required. It would be interesting for future studies to determine why 29% of our engineering students and graduates have chosen to pursue a career outside of space.

We find that 41% of the physics-related respondents have worked in the space sector, now or in the past, compared to only 21% of engineering respondents. Also, of the computer-science and maths-related sample population, a similar proportion of each had at some point worked in the space sector as the proportion of their samples that stated they were intending on one day working in space (17% and 18% respectively). This indicates that SSUK alumni with STEM backgrounds are well represented in the space sector.

The statistics for what area of the space sector respondents work or worked in (only answered by those who do already work in space) show a strong preference for academia, education and research - accounting for over 46% of the total respondents in the space sector, whilst 26% of our alumni working in the space sector work in industry. 16% of respondents identified as working in space education, 25% of whom are mentors.

*2) Influence on Career Choices by Gender*

From Figure 5 we can see that of those alumni currently or previously in the space workforce, the gender split is roughly even. We also find that 25% of male respondents currently work in the space sector compared to 22% of female respondents. In addition, 8% of male respondents stated they had previously worked in the space sector compared to 6% of female respondents. These figures show that, of our alumni in the space sector, women make up 46%. This contrasts, for example, with the UK STEM sector which is 22% women, suggesting that SSUK has a positive influence on gender balance in the UK space sector [2].

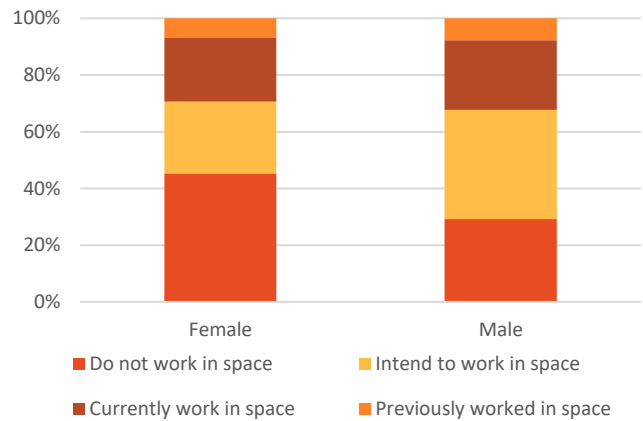

Fig. 5. Answers to the question "Do you currently, or have you in the past, worked in the space sector?" displayed by gender

There is a higher proportion of male respondents intending to work in the space sector than female respondents — 38% and 24% respectively. More than double the number of female respondents than male respondents stated they intended to start a career in STEM however, but not necessarily one in the space sector (nearly 23% of female respondents defined themselves as this vs 11% of male respondents). We would therefore expect a greater representation of female SSUK alumni in STEM fields in the future.

When asked about SSUK's influence on them pursuing a space or STEM career the average female respondents' score is higher than for male respondents (see Figure 6). When separated into their HE subject-related groups this effect is also seen, with each subject group's female respondents registering a higher average than their male counterparts, apart from in maths-related subject (see Figure 7). This reinforces the previous suggestion that SSUK has a greater influence on female attendees.

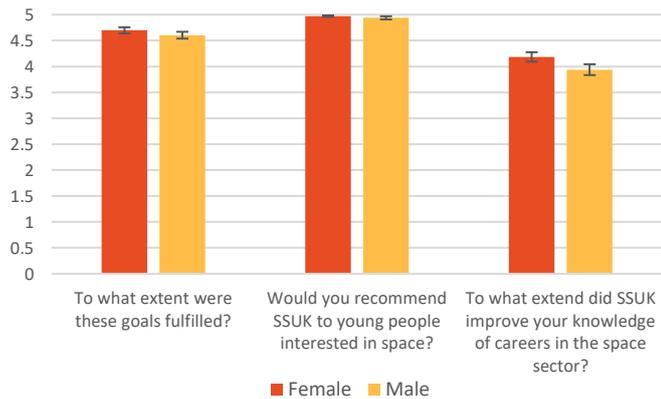

Fig. 6. Average scores of the effect of SSUK on space careers/goals displayed by gender

*3) Career history and overlap between space and STEM*

Responses to our survey questions on work history suggest that SSUK alumni are likely to pursue a career in STEM at some point, with 53% of respondents currently working in STEM or having done so in the past, and an additional 27% intending to do so in the future. Furthermore, all SSUK alumni working in the space sector identify themselves as working in STEM (Figure 8). An additional 25% who have previously worked in the space sector also identified themselves as having worked in STEM. Of those intending to work in STEM in the future, 69% also intend to work in space and 33% of those who have previously worked in STEM still intend to work in space in the future. Overall 60% of respondent alumni stated they intended to or had at some point worked in the space sector.

Figure 8 also seems to indicate that whilst SSUK alumni tend to go into STEM roles within the space sector (and are arguably well supported and encouraged to do so by SSUK), very few of them seem interested in pursuing space from a non-STEM career perspective. As the skills section of this report shows, SSUK is extremely beneficial for those taking up roles in the space sector (as self-professed by respondents), and particularly so for groups that traditionally face added barriers in STEM careers such as women and non-selectively schooled students.

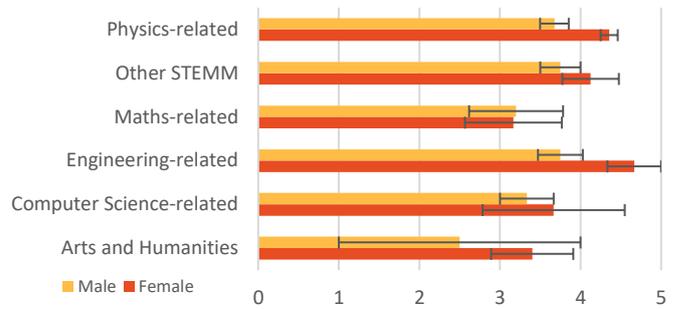

Fig. 7. Average score of answers to 'To what extent did Space School influence you to consider a STEM career?' displayed by gender

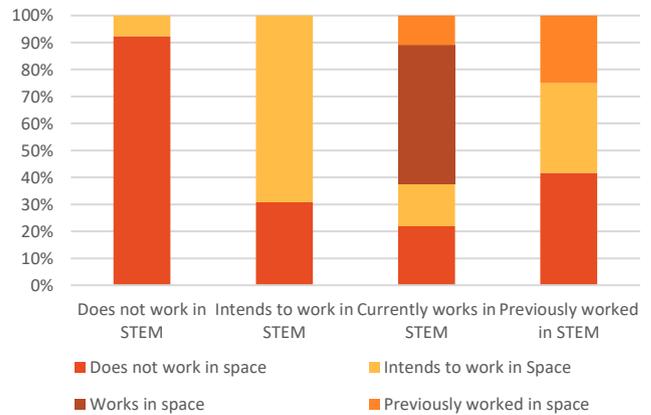

Fig. 8. Answers to "Do you currently, or have you in the past, worked in the space sector?" displayed with answers to "Do you currently, or have you in the past, worked in STEM?"

*4) Investigating influences on career choices by schooling*

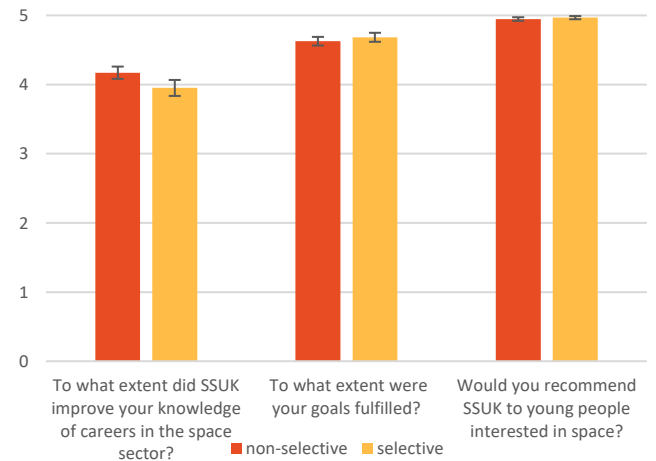

Fig. 9. Average scores to questions on career choices and goals in attending SSUK displayed by schooling type (selective schooling vs non-selective, ie state maintained comprehensive schooling.

When the average scores from pupils from non-selective schooling are compared to those from selective schooling (Figure 9) we can see that both alumni groups would recommend SSUK highly (4.95 ± 0.1 vs 4.97 ± 0.1 respectively). Non-selectively schooled students reported learning more about careers in the space industry than selectively schooled alumni, although selectively schooled alumni polled a slightly higher mean of having their original goals for attending SSUK fulfilled (although they do agree within error). 78% of total respondents answered either a 4 or 5 for the question regarding the improvement of knowledge of space careers - a mean of 4.08 ± 0.1 across all responding alumni.

*5) Outreach and Education*

We find that 75% of respondents who work, or have previously worked, in STEM said that SSUK encouraged them to participate in outreach/education for future generations, whilst a further 15% of those in STEM said it may have done. For those who work/have previously worked in space, the response was largely the same. It is encouraging to also find that 39% of respondents not working in STEM also reported that SSUK encouraged them to participate in outreach/education. Especially as a further 19% of this group said that SSUK may have encouraged them to do so.

However, these data will be skewed, since 28% of those who said that SSUK encouraged them to do outreach/education are mentors. We find that the mentors were the group that rated this question the highest, with 89% of them answering yes to 'Did Space School encourage you to participate in Outreach/Education for Future Generations', which indicates that SSUK encouraged this group enough to pursue outreach/education by becoming a mentor.

### D. Impact of SSUK on skills development

As well as their views on HE and careers, alumni were also asked questions about their goals of attending SSUK and the skills they felt they took away from it.

When asked what their goals were in attending SSUK, where they could tick all that apply, the main reasons respondents cited were to learn more about space/astronomy (93%), meet like-minded people (68%) and to learn more about possible space careers (51%). When asked what skills they learnt by attending SSUK, knowledge and understanding of space was highest (87%), followed by social confidence (83%) and teamwork (71%).

We find that SSUK has a strong positive effect on young people's 'soft skills' and that these were cited as being the most beneficial when asked which category has helped them the most with your studies/career. Socialising/building a space network was cited the most, at almost 25% (see Figure 10). This percentage is higher for female respondents than male respondents (see Figure 11), suggesting the positive influence of SSUK on gender balance, as, for example, the gender balance in SSUK (~40% female historically) is higher than the gender balance of A-level physics (~20 %, according to [3]). This higher representation of women may contribute towards the higher scores given by female alumni in this survey. This result is important, as it clearly demonstrates the extra value that providing a positive environment and role models for girls is key in encouraging them to pursue careers in STEM and the space sector, as many alumni have done.

Furthermore, female respondents reported that SSUK increased their skills in teamwork and problem solving more than male respondents, whilst male respondents cited increased skills in communication, organization, practical and technical skills over female respondents.

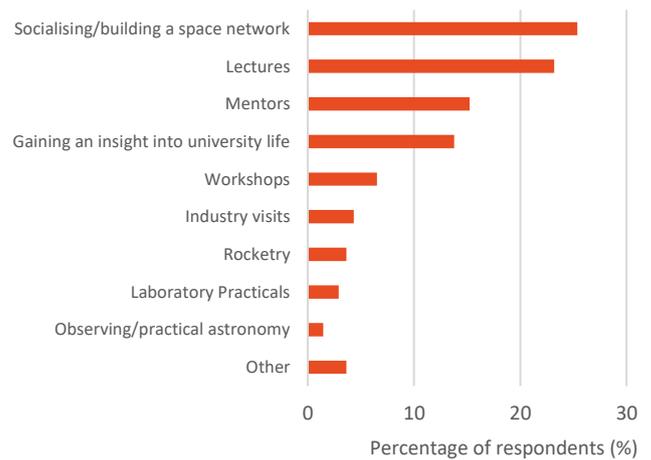

Fig. 10. Most common answers to 'Which of these categories would you say has helped you most with your studies/career?

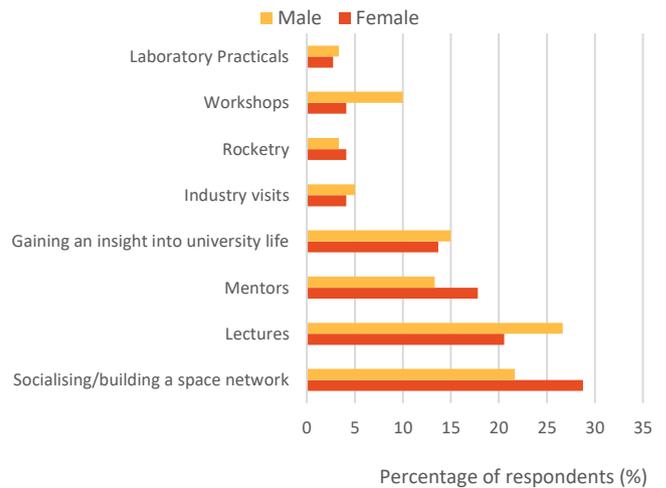

Fig. 11. Percentage of male and female respondents' answers to 'Which of these categories would you say has helped you most with your studies/career?'

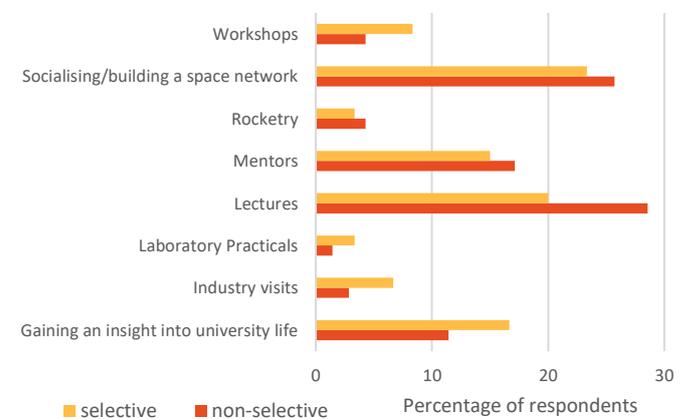

Fig. 12. Percentage of selectively and non-selectively schooled respondents' answers to 'Which of these categories would you say has helped you most with your studies/career?'

SSUK not only improves confidence and teamworking skills, but 44% of those who have gone on to work in the space sector have identified that socialising and building a space network was the most useful aspect of the summer school.

When these skills-based questions were grouped by schooling, Figure 12 was produced showing alumni's self-reported most beneficial aspect of SSUK to their career. Some notable points from this are that selectively schooled students found workshops, industry visits and getting an insight into university life to be the most useful aspects of SSUK. Conversely, non-selectively schooled students found the lectures significantly more useful, and mentors slightly more so.

## IV. Limitations of this study

Due to time constraints, the survey was only available for two weeks via social media, reducing the potential reach. Due to GDPR constraints, i.e. lack of alumni email addresses, social media and word of mouth were the main forms of sharing the survey. Although SSUK social media groups have existed since 2007, members are predominantly more recent alumni, younger people, and current mentors, so were more likely to have completed the survey over people who attended SSUK longer ago. Over half (52%) of respondents attended SSUK recently (2014-2018 inclusive) at least once, and 71% of respondents had attended between 2009 and 2018. This therefore skews the dataset away from those in their mid-late careers.

Of those aware and with access to the survey, those who are more engaged with the SSUK community and space in general are more likely to have been motivated to complete the survey than those less enthused. This will systematically skew some data. 40% of respondents attended SSUK once, whilst 60% reported attending SSUK twice or more. 19% of respondents work, or have previously worked, as a mentor at SSUK. These mentors are especially likely to perceive SSUK in a more positive light and with a greater influence on their lives, as are those that came multiple times as they are presumed to have enjoyed it enough to attend again.

Many alumni may have already had the resources, skills, connections and interest to pursue HE and/or a career without attending SSUK. However, this work shows SSUK has still reinforced interest in space and broadened students' views of the space sector and career choices in nearly all respondent cases.

## V. Conclusions

SSUK teaches skills and provides opportunities that many respondents feel were critical for them to study and establish careers within the STEM and space industry. Some groups reported stronger than average positive influences from SSUK; these include women and students from non-selective school backgrounds. We find this is particularly the case for alumni from these groups entering HE, for example, both the female and non-selectively schooled alumni's average score for 'What impact did SSUK have on your opinions/plans of entering HE' was 4.5, higher than their male and selectively schooled counterparts (4.2 and 4.3 respectively).

The survey finds that SSUK primarily increased knowledge and understanding of space, but also increases students 'soft skills', such as teamwork and communication. SSUK adds value, regardless of background, to aspirations for careers and higher education, building technical and soft skills.

We find that a large proportion of respondents have at some point worked in STEM (53%) and 30% have also worked in the space sector, meaning that alumni are well represented in these areas in the UK. Another 27% of respondents indicate they have the intention to work in STEM in the future, with 69% of these also intending to join the space sector. Furthermore, SSUK tends to encourage alumni to participate in outreach/education themselves, which is an important part of continuing SSUK's legacy.

As a result of this survey, the authors encourage our likeminded colleagues to provide networking opportunities in space for young people. 43.8% of respondents who have gone on to work in the space sector have identified that socialising and building a space network was the most useful aspect of the summer school. SSUK generates long lasting impacts for its participants, particularly women, through creating and maintaining a space network from a young age.

The UK Space Agency's goal for the UK to make up 10% of the global space sector by 2030 will only be reached by increasing the size of the workforce [4]. SSUK inspires young people to join the sector through giving insights into work and study in the field. Through this work we have shown that SSUK is in a unique position to provide a space careers pipeline and network in the UK.

Acknowledgements

The authors acknowledge the School of Physics and Astronomy at the University of Leicester for hosting Space School UK and Senior Space School UK. The authors gratefully acknowledge current and previous sponsors and supporters, including and not limited to the UK Space Agency, Lockheed Martin UK and The Air League, for their support of Space School UK and Senior Space School UK and provision of scholarships for students. The authors thank the Space School UK and Senior Space School UK alumni for partaking in this 30th anniversary survey.